\documentclass[letter]{aa}
\usepackage{txfonts}
\usepackage{natbib}
\bibpunct{(}{)}{;}{a}{}{,} 
\usepackage{graphicx}

\newcommand{\be}{\begin{equation}}
\newcommand{\ee}{\end{equation}}

\newcommand{\sig}{\left<\sigma v\right>}
\newcommand{\rhos}{\rho_{\rm s}}
\newcommand{\rs}{r_{\rm s}}
\newcommand{\dm}{_{\rm dm}}
\newcommand{\g}{_{\rm gas}}
\newcommand{\fa}{f_{\rm abs}}
\newcommand{\fr}{f_{\rm rad}}
\newcommand{\tc}{\tau_{\rm c}}
\newcommand{\tdyn}{\tau_{\rm dyn}}

\newcommand{\Esn}{\epsilon_{\rm SN}}
\newcommand{\rb}{r_{\rm b}}

\newcommand{\msun}{{\rm M}_\odot}

\begin{document}

\title{Effect of dark matter annihilation on gas cooling and star formation}

\date{\today}

\author{Yago Ascasibar}
\institute{Astrophysikalisches Institut Potsdam, An der Sternwarte 16, Potsdam D-14482 (Germany)\\ \email{yago@aip.de}}


\abstract
{In the current paradigm of cosmic structure formation, dark matter plays a key role on the formation and evolution of galaxies through its gravitational influence.
On microscopic scales, dark matter particles are expected to annihilate amongst themselves into different products, with some fraction of the energy being transferred to the baryonic component.}
{It is the aim of the present work to show that, in the innermost regions of dark matter halos, heating by dark matter annihilation may be comparable to the cooling rate of the gas.}
{We use analytical models of the dark matter and gas distributions in order to estimate the heating and cooling rates, as well as the energy available from supernova explosions.}
{
Depending on the model parameters and the precise nature of dark matter particles, the injected energy may be enough to balance radiative cooling in the cores of galaxy clusters.
On galactic scales, it would inhibit star formation more efficiently than supernova feedback.
}
{Our results suggest that dark matter annihilation prevents gas cooling and star formation within at least $0.01-1$ per cent of the virial radius.}


\maketitle

\section{Introduction}
\label{secIntro}

One of the most remarkable achievements of modern Cosmology is the measurement of the fundamental constituents of the Universe.
Over 80 per cent \citep{Spergel+_06} of the matter (one fifth of the total energy density) is currently believed to be composed of non-baryonic dark matter particles \citep[see e.g.][for a review of candidates]{Bertone+05}.
The cold dark matter scenario has been extremely successful in explaining many of the observed properties of galaxies over a broad range of scales and environments, but there are nevertheless several issues that still defy our understanding.

One of them is the number of dwarf galaxies orbiting around the Milky Way, with numerical simulations predicting one or two orders of magnitude more satellites than observed \citep{Klypin99,Moore99}.
Perhaps the currently most favored explanation is that photoionization \citep{Efstathiou92} prevented gas cooling in the smallest objects.
According to \citet{Kravtsov04}, star formation should be strongly suppressed for all halos smaller than $10^9~\msun$.
A similar problem might exist on galaxy cluster scales, which would push the threshold to $10^{11}~\msun$ or even larger \citep{Kase_06}.

Actually, observations of the conditional luminosity function indicate that the mass-to-light ratio reaches a minimum for halo masses around $3\times10^{11}~\msun$, with objects below $10^{10}~\msun$ virtually devoid of galaxies \citep{Bosch03}.
It is at present unclear whether the cosmic ultraviolet background could provide enough photons to achieve such effect; numerical experiments suggest that some fraction of the gas is still expected to cool and collapse into stars at the centers of most halos \citep{Hoeft_05}.

There is an observed upper mass limit above which star formation seems to be suppressed as well.
This threshold increases with redshift \citep[see e.g.][]{Bundy+06}, in apparent contradiction with the hierarchical picture.
Star formation in the most massive galaxies takes place at relatively early times and then it suddenly shuts off, while less massive objects tend to form their stars at later times \citep{Cowie96}.

The problem is particularly noticeable in galaxy clusters, where it is difficult to understand why the gas is currently cooling at a much slower rate than expected from its X-ray luminosity \citep{Peterson01}, and an external heat source, such as an active galactic nucleus, is often invoked in order to explain the phenomenon.

A similar mechanism may also be responsible for the red and blue sequences observed in the color-magnitude diagram of galaxies, that can only be reproduced by quenching star formation in halos more massive than $\sim10^{12}~\msun$ \citep{Croton06,Cattaneo_06}.

Here we propose that annihilation of dark matter particles may provide a considerable amount of energy, which, if transferred to the surrounding gas, could help alleviating some of the discrepancies outlined above.

On cluster scales, it has been argued that neutralino annihilation may play a role in the cooling flow problem \citep{Totani04,Colafrancesco+06}, and there have been several studies assessing the impact of dark matter decay and/or annihilation on the cosmic ionization history \citep{PadmanabhanFinkbeiner05,Mapelli+06,Zhang_06} and the soft gamma-ray background \citep{AhnKomatsu05,Rasera+06}, as well as on the first generation of galaxies \citep{Ripamonti+_06_gal}.

It is our aim to show that, for a relatively broad range of scenarios, dark matter annihilation may also influence galaxy formation and evolution by quenching gas cooling and star formation near the center of dark matter halos.

\section{Heat from cold dark matter}
\label{secHeat}

When two dark matter particles annihilate, all their energy goes into the annihilation products.
Some fraction, $\fr$, will be radiated away at different wavelengths, from gamma rays to radio, and it can be used to impose constraints on the nature of dark matter \citep{Bertone+05}.
The present study focuses on the remaining fraction, $\fa=1-\fr$, that is eventually absorbed by the surrounding baryonic gas, which is thus heated at a rate
\be
\dot u
= \fa\,n\dm n_{\rm dm*}\sig\,2m\dm c^2
= \fa\,C\rho\dm^2 \sig c^2 / \,(2m\dm)
\label{eqE}
\ee
where the dot denotes derivative with respect to time, $u$ is the gas energy per unit volume, $m\dm$, $\rho\dm$, and $n\dm$ are the mass, density, and number density of dark matter particles, respectively, $\sig$ is their annihilation cross-section, and we have assumed in the last step non-self-conjugate particles, $n\dm=n_{\rm dm*}=\rho\dm/(2m\dm)$.

The clumping factor, $C=\left<\rho\dm^2\right>/\left<\rho\dm\right>$, accounts for the presence of substructures \citep{Bergstroem+99}.
Current simulations \citep{Diemand+_06} indicate that $C$ should be of the order of a factor of 2 or 3, although higher values \citep[e.g.][]{Colafrancesco+06}, $C\ge10$, cannot be excluded.
It is important to note, though, that the distribution of subhalos is less concentrated than that of the smooth component of the main halo \citep{NagaiKravtsov05}, and therefore $C$ should actually be a decreasing function of $\rho\dm$.

It has been recently shown that, for light dark matter particles ($m\dm\sim3-10$~MeV) annihilating in a neutral unmagnetized gas at the average cosmic density, $\fa\ge0.03$ \citep{Ripamonti+_06_fabs}.
At typical galactic densities, the absorbed fraction is expected to increase due to the higher rate of Coulomb collisions, up to $\fa\sim1$.
However, for heavier candidates ($m\dm>100$~MeV) more radiation is expected through inverse Compton and synchrotron emission from the initially relativistic annihilation products.
Thermalization becomes again efficient when the energy of these particles (most notably electrons and positrons, but also protons and other particles) drops below a few GeV, leading to $\fa\sim0.1$ for $m\dm<100$~GeV \citep[][]{Totani04}.

For thermal relics, the observed cosmic density imposes $\sig\simeq3\times10^{-26}$~cm$^3$~s$^{-1}$ at the time of decoupling.
However, light particles with $0.511<m\dm<100$~MeV and such a cross-section would produce too many positrons and gamma rays within our galaxy.
Compatibility with INTEGRAL/SPI measurements requires \citep{Ascasibar_05} that the present-day annihilation cross-section into electron--positron pairs satisfies
\be
\sig_{e^+ e^-} \leq 3\times10^{-30}\left(\frac{m\dm c^2}{\rm 1~MeV}\right)^{2}~{\rm cm^3\ s^{-1}}.
\label{eqSig}
\ee
Substituting this expression in equation~(\ref{eqE}), the energy input increases proportionally to $m\dm$ until it reaches a maximum at $m\dm\sim100$~MeV, and then it declines as $m\dm^{-1}$.

For the density profile of the dark matter halo, we adopt the general formula
\be
\rho\dm=\frac{\rhos}{(r/\rs)^{\alpha}\,(1+r/\rs)^{3-\alpha}}
\label{eqRhoDM}
\ee
where $\rhos$ and $\rs$ are the characteristic density and radius of the object, and $\alpha$ is the asymptotic logarithmic slope at the center.
Cosmological N-body simulations \citep{NFW97} suggest $\alpha\sim1$.
Lower values have been inferred from the rotation curves of dwarf and low surface brightness galaxies \citep{FloresPrimack94,Moore94}, although it is at present unclear whether these observations may be consistent with steeper profiles \citep{Hayashi+04,Spekkens+05}.
On the other hand, $\alpha>1$ is expected from adiabatic contraction due to stars \citep{Blumenthal+86} and/or a supermassive black hole \citep{GondoloSilk99}.
Finally, the density profile
\be
\rho\dm=\rho_2\, e^{ -\frac{2}{\beta} \left[ (\frac{r}{r_2})^\beta -1 \right] }
\label{eqN04}
\ee
with $\beta\simeq0.18$ also describes the results of numerical simulations \citep{Navarro+04}, and it tends to a finite value at the center.
In this expression, $\rho_2$ is the density at the radius $r_2$ where the logarithmic slope is equal to $-2$.
For the profile~(\ref{eqRhoDM}), this occurs at $r_2=(2-\alpha)\rs$.
In order to model halos on different scales, we account for the mass-concentration relation \citep{Bullock+01} according to
\be
\frac{r_2}{20~{\rm kpc}}= \left( \frac{M_{200}}{10^{12}~\msun} \right)^{0.46}
\ee
where $M_{200}$ is the mass enclosed at an overdensity of 200 times the critical density.

\section{Results}
\label{secResults}

\begin{figure}
  \includegraphics[width=9cm]{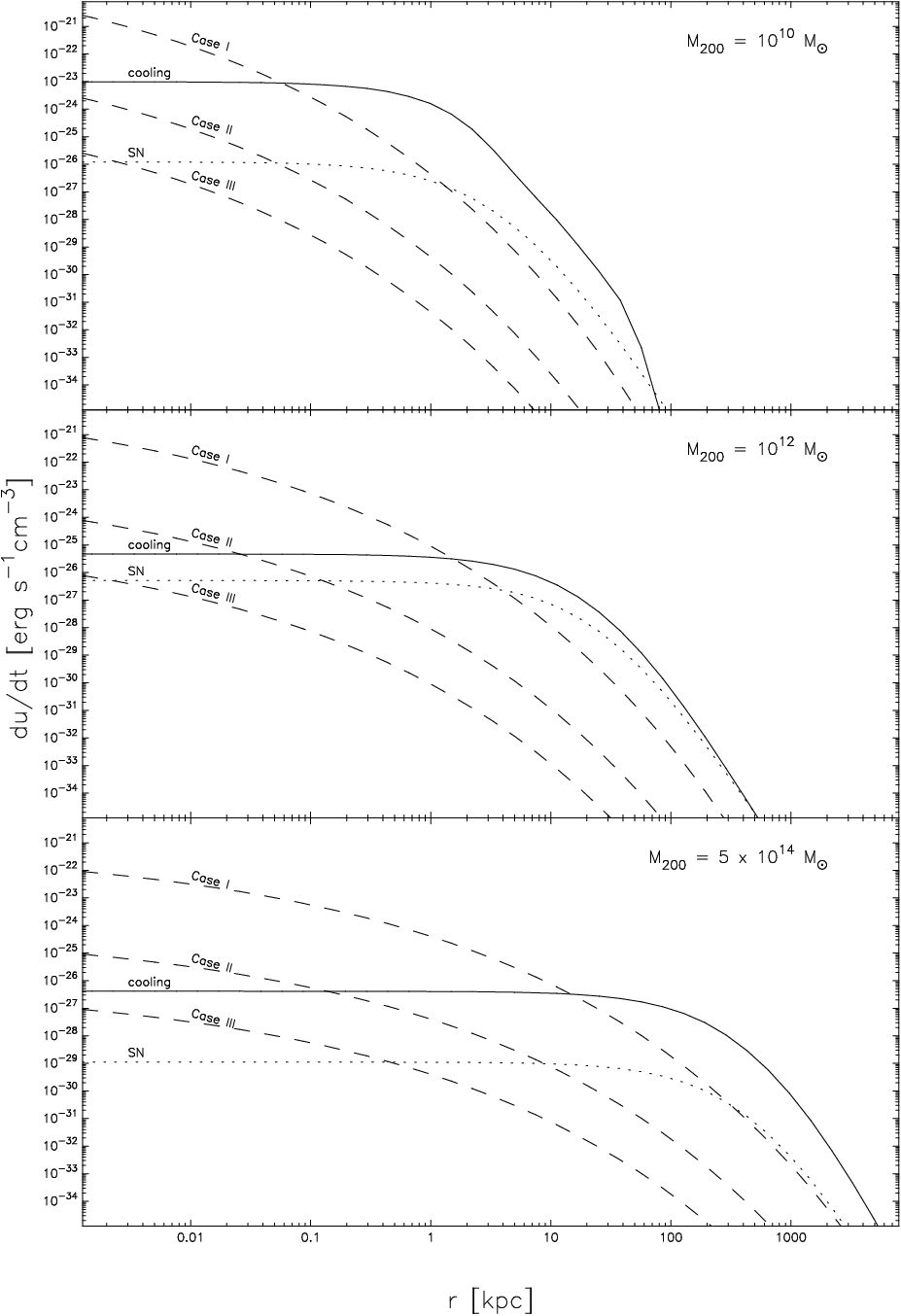}
  \caption
  {
    Heat injection (dashed lines) for different dark matter scenarios (see text), compared to the cooling rate of the gas (solid lines) and feedback from supernovae (dotted lines) in halos of different mass.
  }
  \label{figUdot}
\end{figure}

\begin{figure}
  \includegraphics[width=9cm]{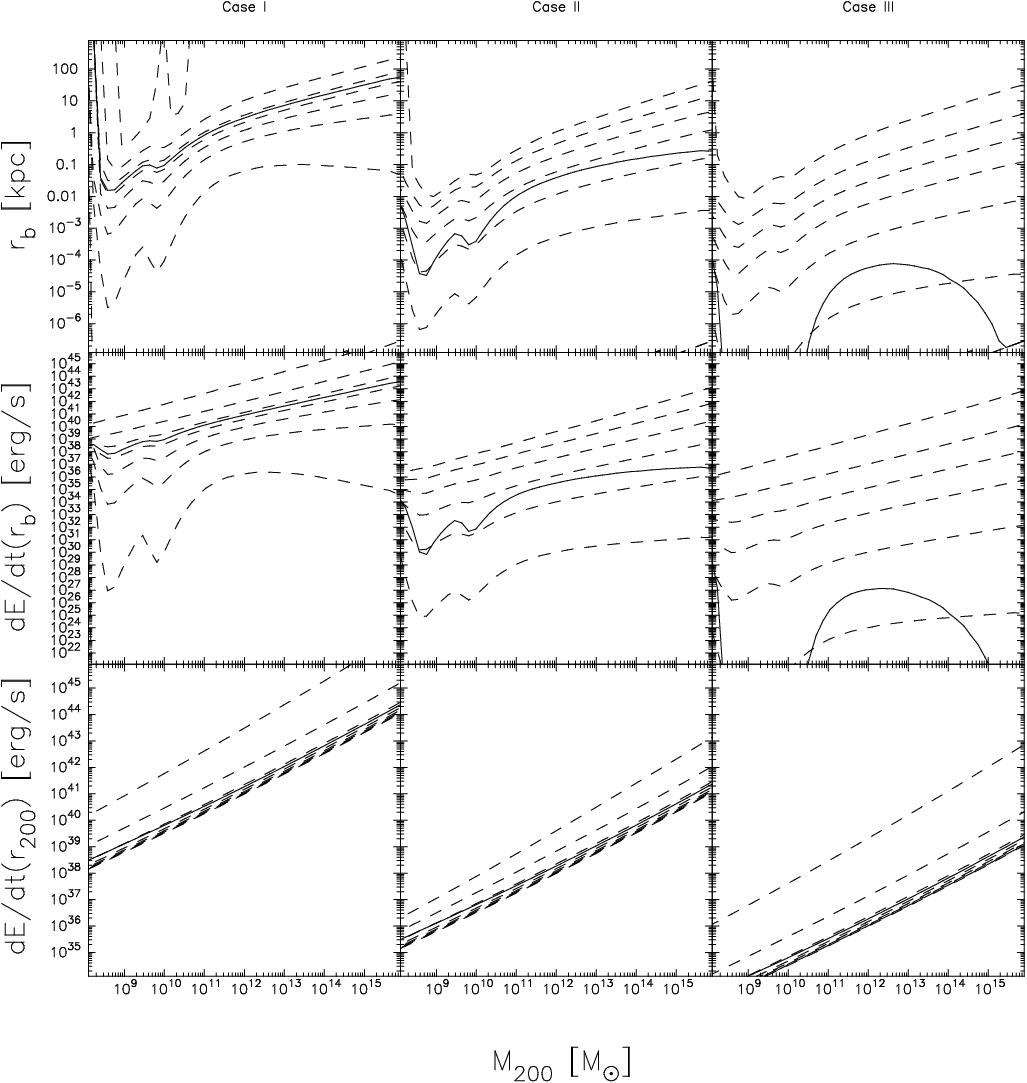}
  \caption
  {
    Radius $\rb$ at which heating balances cooling (top), energy injection within $\rb$ (middle) and within $r_{200}$ (bottom), as a function of virial mass.
    Solid lines correspond to the profile~(\ref{eqN04}).
    Results for expression~(\ref{eqRhoDM}), with $\alpha$ varying from 0.25 to 1.75 in steps of 0.25, are plotted as dashed lines.
  }
  \label{figRhoDM}
\end{figure}

We compare in Figure~\ref{figUdot} the heating rate given by equation~(\ref{eqE}) with the local cooling rate of the gas and the energy available from supernova feedback.
In order to obtain an upper limit (Case I), we set $m\dm=100$~MeV, $\sig=3\times10^{-26}$~cm$^3$~s$^{-1}$, $\fa=1$, and $C=10$.
More realistic values, $\fa=0.1$ and $C=1$, are assumed for Case II ($m\dm=10$~MeV, $\sig=3\times10^{-28}$~cm$^3$~s$^{-1}$) and Case III ($m\dm=100$~GeV, $\sig=3\times10^{-26}$~cm$^3$~s$^{-1}$).

The dark matter density profile follows expression~(\ref{eqN04}), and the baryonic component is modeled as a polytrope with effective index $\gamma\simeq1.18$ \citep{Ascasibar03}.
The cooling rate of the gas is determined by the function $\Lambda(T\g)=\dot u/\rho\g^2$ tabulated in \citet{SD93}, and new stars are formed as \citep{Kennicutt98}
\be
\dot\rho_* \simeq 0.02\,\frac{\rho\g}{\tdyn}
\ee
where $\tdyn=\sqrt{(3\pi)/(16G\rho\g)}$ is the local dynamical time\footnote{
If the cooling time $\tc=u/\dot u<\tdyn$, the gas can cool to temperatures $\sim10^4$~K in pressure equilibrium with the ambient medium, and its density will be enhanced by a factor $T\g/(10^4~{\rm K})$.
}.
Supernova explosions can inject energy at a rate $\dot u = \Esn\dot\rho_*$, with $\Esn\simeq4\times10^{48}$~erg~$\msun^{-1}$ for a \citet{Salpeter55} initial mass function.

Heat from annihilating dark matter particles is usually several orders of magnitude below the radiative cooling rate of the gas, except in the central part, where the dark matter density becomes much larger than the gas density.
In this region, dark matter annihilation not only provides more energy than supernovae; in fact, it would prevent gas cooling and star formation completely.

The heated gas would expand and rise buoyantly, creating winds, shocks, and turbulence, and a full three-dimensional simulation would be required in order to evaluate the net effect on the ambient baryonic medium.
Recent numerical studies of the effect of cosmic rays accelerated in structure formation shocks show that the total mass-to-light ratio of small halos and the faint end of the luminosity function can indeed be strongly affected by the injection of relativistic particles \citep{Jubelgas+_06}.

The extent of the heating-dominated region is very sensitive to the specific dark matter candidate, which sets the normalization of equation~(\ref{eqE}) through the product of $\fa$, $\sig$, and $m\dm$.
We show in Figure~\ref{figRhoDM} the radius $\rb$ at which heating balances cooling, as well as the integrated energy injection within $\rb$ and $r_{200}$, for cases I, II, and III.
We also plot the results for a dark matter density profile of the form~(\ref{eqRhoDM}) for $0.25<\alpha<1.75$.

When $\alpha=0$, dark matter annihilation is not able to counteract gas cooling at any radius, not even in the most optimistic Case I.
However, this does not apply to any ``cored'' density profile.
Our results for expression~(\ref{eqN04}) are similar to those obtained for $\alpha=1$ in cases I and II.
In Case III, dark matter heating is so close to gas cooling that the asymptotic behaviour of the density profile has a critical impact on $\rb$.

\begin{figure}
  \includegraphics[width=9cm]{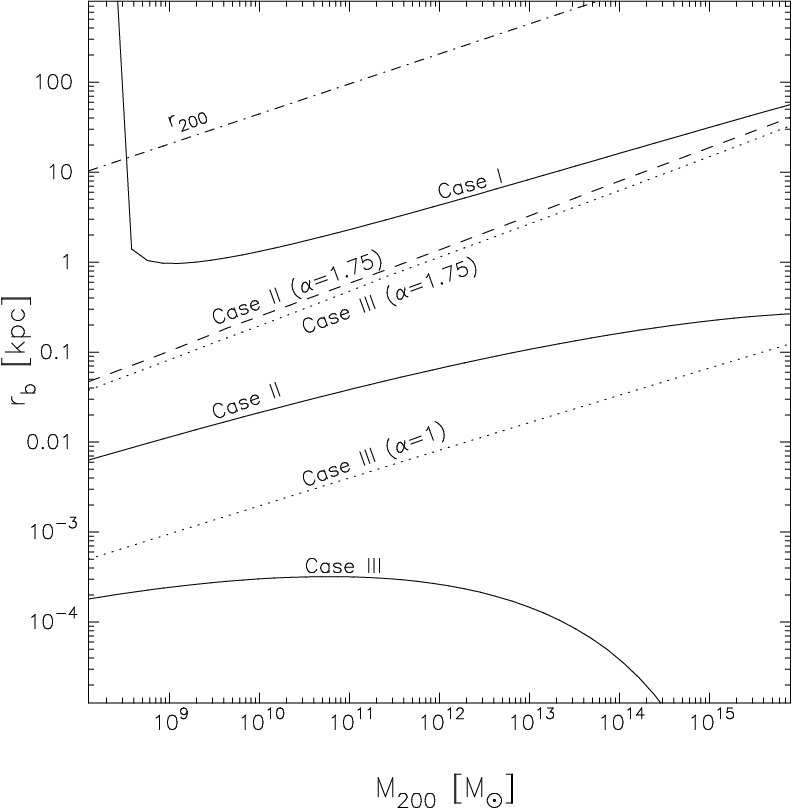}
  \caption
  {
    Radius $\rb$ for an ionized gas.
    Solid lines depict cases I, II, and III (see text).
    Dashed line corresponds to Case II with $\alpha=1.75$ in~(\ref{eqRhoDM}), and dotted lines show $\alpha=1$ and $\alpha=1.75$ for Case III.
    Dash-dotted line displays the virial radius, $r_{200}$.
  }
  \label{figIon}
\end{figure}

Values of $\rb$ range from sub-parsec to kpc scales.
For the values of $C$ and $\fa$ adopted in cases II and III, dark matter annihilation could only solve the cooling flow problem for a steep ($\alpha\ge1.75$) profile, in agreement with previous studies \citep{Totani04,Colafrancesco+06}.
On galactic scales, it will reduce the fraction of baryons collapsed into stars and prevent the growth of excessively massive bulges in small systems.
Ionization
\footnote
{
The time required to ionize all gas within $\rb$, $\tau_{\rm ion}\sim M\g(\rb)\left<E_{\rm ion}\right>/\dot E(\rb)$, where $\left<E_{\rm ion}\right>\simeq3\times10^{46}$~erg~$\msun^{-1}$, is of the order of 1 Myr.
}
would reduce the cooling rate for gas temperatures $10^4<T\le10^5$~K and increase it for $T<10^4$~K.
Figure~\ref{figIon} gives an estimate of this effect by assuming pure thermal bremsstrahlung \citep{Efstathiou92}.
Only the unrealistic Case I would be able to completely quench star formation in small ($M_{200}<10^9~\msun$) halos.
In all the other cases, $\rb\le0.01r_{200}$.

Finally, we have tested the dependence on the shape of the gas profile by using equation~(\ref{eqRhoDM}) to model the baryonic component.
Results depend very weakly on the value of $\alpha\g$, as long as $\alpha\g\le0.3$.
For higher values, which may be attained if cooling has already taken place in the outer parts of the halo, dark matter annihilation will not be effective in preventing further cooling and star formation.
Note, however, that this gas will be quickly converted into stars, leaving a less dense medium.
  In fact, such mechanism has been proposed to explain the entropy excess in galaxy groups \citep{Bryan00}.

\section{Conclusions}
\label{secConclus}

Annihilation of dark matter particles would produce gamma ray \citep[see e.g.][and references therein]{Bertone_06}, X-ray \citep[e.g.][]{Bergstroem+_06} and radio \citep[e.g.][]{ColafrancescoMele01,Colafrancesco04} emission.
The present study suggests that it may also have a noticeable effect on galaxy formation and evolution, providing a constant (and relatively powerful) heat source at the center of every dark matter halo.

The magnitude of the effect depends on the physical properties of dark matter and its distribution within the halo.
Our analytical estimates show that, for reasonable values of the model parameters, the amount of energy injected into the gas can be larger than the cooling rate in the central regions.
Only for extremely shallow dark matter density profiles or steep gas density profiles could all the energy be radiated efficiently.

Else, cooling and star formation would be completely switched off within the radius $\rb$ where heating balances cooling.
For most of the cases considered, this radius is between 0.01 and 1 per cent of the virial radius of the object.
For the upper limit, Case I, our results indicate that no stars could form within 10 per cent of the virial radius, and no star at all could form in haloes less massive than $\sim10^9~\msun$.

Evaluating the impact of dark matter annihilation on the star formation rate outside $\rb$ would only be possible by implementing dark matter heating in a self-consistent numerical simulation of cosmic structure formation.
Equation~(\ref{eqE}) provides a simple prescription to carry out such an experiment and determine the maximum amount of heat compatible with current observations.
Some dark matter scenarios (e.g. Case I) seem to inject too much energy for galaxies to form.
More realistic models (e.g. cases II and III) might actually explain why current models predict many more stars than observed, particularly in the central regions of galaxies, thus providing an intriguing alternative to more conventional astrophysical feedback mechanisms.

\begin{acknowledgements}
The author would like to thank C\'eline B\oe hm for many useful comments and discussions, without which this work would have never been possible.
\end{acknowledgements}

\bibliographystyle{aa}
\bibliography{heat}

\end{document}